\documentstyle[aasms4]{article}

\makeatletter
\def\eqalign#1{\null\,\vcenter{\openup\jot\m@th
  \ialign{\strut\hfil$\displaystyle{##}$&$\displaystyle{{}##}$\hfil
      \crcr#1\crcr}}\,}
\def\eqalignleft#1{\null\,\vcenter{\openup\jot\m@th
  \ialign{\strut$\displaystyle{##}$\hfil&$\displaystyle{{}##}$\hfil
      \crcr#1\crcr}}\,}
\makeatother
\def\M{{\cal M}}
\def\R{{\cal R} }
\lefthead{Postnov and Prokhorov}
\righthead{Binary GW Noise}
\begin{document}

\title{Galactic Binary Gravitational Wave Noise within LISA 
Frequency Band}

\author{Konstantin A.~Postnov\altaffilmark{1}
and Mikhail E.~Prokhorov\altaffilmark{2}}

\affil{Sternberg Astronomical Institute, Moscow State University, 
119899, Moscow, Russia}
\altaffiltext{1}{Visiting Scientist,
The Institute of Physical and Chemical Research (RIKEN), 2-1 Hirosawa,
Wako, Saitama, 351-01, Japan
\hbox{e-mail}: pk@sai.msu.su}

\altaffiltext{2}{e-mail: mike@sai.msu.su}
\date{Received  1997 January 29; accepted ..., 1997}

\begin{abstract}

Gravitational wave noise associated with unresolved binary stars in
 the Galaxy is studied with the special aim of determining the upper
 frequency at which it stops to contribute at the rms noise level of
 the proposed space-born interferometer (LISA).  The upper limit to
 this background is derived from the statistics of SN Ia explosions,
 part of which can be triggered by binary white dwarf coalescences.
 The upper limiting frequency at which binary stochastic noise crosses
 LISA rms sensitivity is found to lie within the range $\approx
 0.03-0.07$ Hz, depending on the galactic binary white dwarf
 coalescence rate. To be reliably detectable by LISA, the energy
 density of relic cosmological background per logarithmic frequency
 interval should be $\Omega_{GW}h_{100}^2>10^{-8}$ at $f>0.03$ Hz.

\keywords{Binaries: close -- gravitation -- waves -- white dwarfs}

\end{abstract}

\section{Introduction}

Binary systems constitute at least half stellar galactic population
($\sim 10^{11}$ stars) and are reliable sources of gravitational waves
(GW). For a binary system consisting of two solar-mass stars, the
characteristic time of orbital decay due to angular momentum removal
by GW becomes shorther than the Hubble time ($\approx 15\times 10^9$
years) if its orbital period is less than about 14 hours. Observations
of binary pulsars with a neutron star (NS) as secondary component
(Taylor 1992) provides us with the strongest observational evidence
for this fundamental process, and the most compact NS or black holes
(BH) binaries merging due to GW emission are considered as primary
real targets for the initial GW iterferometric LIGO-type detectors
(Abramovici et al. 1992).  An enormous energy is released during the
coalescence of a compact NS+NS binary (typically, of the order of
$10^{53}$ ergs), but the galactic merging rate of NS+NS binaries is
fairly small, $10^{-4}$ -- a few $10^{-5}$ per year (see Lipunov,
Postnov \& Prokhorov 1997 for more detail), and to detect an
acceptable number of such mergers per year (the experimentalists
usually quote the number of 3 events per year with the signal-to-noise
ratio 3-5), they are planned to be studied from distances up to 200
Mpc (e.g., Thorne 1987). Typical frequencies of GW emitted during
compact binary NS coalescence are about 100-1000 Hz, depending on the
mass of the stars involved.

Orbital frequencies at which a binary system may be observable are
limited by the size of the components; the evolution of the orbital
separation strongly changes when one of the stars fills its critical
Roche lobe. Typically, mass exchange between the components has much
larger effect on the binary separation than the orbital angular
momentum removal due to GW, so the orbital frequency of a given binary
system may fall within a wide range from a fraction of hour to several
years, depending on the initial parameters and details of evolution.
The evolution of any star, however, ends up with the formation of a
compact remnant (BH, NH or white dwarf (WD)), so when two such compact
stars remains in a binary, its orbital evolution is totally controlled
by the removal of orbital angular momentum by GW (the possible rare
exceptions are very hot young WD with a strong stellar wind or the
cases where a low-mass WD companion is evaporated by a strong
pulsar emission; we will not consider them here). When only GW emission
drives the orbital evolution, a simple analytical treatment is
in order.

As the number of binary stars in the Galaxy is very large, the GW
emitted by them (at strictly twice the orbital frequency if the orbit
eccentricity is zero) form a stochastic background in the frequency
range from $10^{-7}$ to $\sim 1$ Hz (Mironovskij, 1965; Rosi and
Zimmerman 1976; Lipunov and Postnov 1987; Lipunov et al. 1987; Hills
et al. 1990; Lipunov et al. 1995).  Being interesting by itself, this
background, however, is viewed as a noise burying a possible
cosmological gravitational wave background (CGWB), which bears the
unique imprint of physical processes occurring at the very early
(near-Plankian) age of the Universe (see e.g. Grishchuk 1988 for a
review).

A stochastic GW background is commonly measured in terms of the energy
density per logarithmic frequency interval related to the critical
energy density to close the Universe, $\Omega_{GW}=dE_{GW}/d\ln
f/\rho_{cr} c^2$ ($\rho_{cr}\approx 1.9\times 10^{-29} h_{100}^2$ g
cm$^{-3}$ where $h_{100}=H_0/100$ km s$^{-1}$ Mpc$^{-1}$ is the
present value of the Hubble constant, $c$ is the speed of light).
Cosmic microwave background fluctuations experimentally detected in
the last years (Smoot 1992; Strukov et al. 1993) put upper bounds on
CGWB of order $\Omega_{GW} = 10^{-14}$ at both LISA and LIGO
frequencies assuming equation of state $p=-\epsilon$ at the
inflationary stage of the Universe.  In this model, CGWB spectrum is
inversely proportional to the frequency keeping
$\Omega_{GW}(f)=constant$ at $f>10^{-14}$ Hz (Rubakov et al. 1982).
If real, this CGWB has no chance to be detected with ongoing GW
detectors (see Schutz 1995, 1996; Allen 1996).  We should note,
however, that the measurements of cosmic microwave background
temperature fluctuations put constraints on the CGWB only at very low
frequencies $f_H\sim H_0\sim 10^{-18}$ Hz, so the dependence of CGWB
energy density upon frequency is very crucial.

The situation, however, starts to change after a recent processing
of the COBE data has provided us with some fresh information about the
power-law spectral index of primordial perturbations (Bennett et
al. 1996; Brukhanov et al. 1996). The point is that the spectrum of the
primordial perturbations may be directly recalculated into the
spectrum of relic GW background (see Grishchuk 1996 and references
therein). According to Grishchuk (1996), the fact that the COBE data
point to some deviations from Harrison-Zel'dovich spectrum of the
initial perturbations translates into the deviations from
$p=-\epsilon$ equation of state at the epoch of inflation which would
lead to serious changes in the CGWB as a whole.  Now its energy
density may increase with frequency reaching $\Omega_{GW}=10^{-8}$ at
$f=10^{-2}$ Hz.  This conclusion holds regardless of the real cause of
the observed CMB temperatrure fluctuations (either they are due to
density fluctuations or relic gravitational waves).  If Grishchuk's
calculations are correct, this opens a novel possibility of detecting
CGWB with the proposed LISA space interferometer (Schutz 1996). The
conventional technique for detection of the stochastic GW backgrounds
assumes the cross-corelation of outputs of at least two independent
interferometers (Grishchuk 1976; Compton \& Schutz 1996 and reference
therein). The LISA project does not plan to have two independent
interferometers, however the strength of the signal
($\Omega_{GW}=10^{-8}$) is predicted so high that probably the signal
could be directly detected.

This important finding raises the question: At which frequency GW
confusion limit from binary systems becomes lower than the
detectability threshold dictated by the signal to noise ratio equal
one? In other terms, beginning at what frequency can we be sure
that no known noise sources of astrophysical origin exist and
therefore if detected, only cosmological background can contribute at
these frequencies?

The aim of the present paper is to answer this question using new
calculations of binary stellar evolution (see Lipunov et al. 1995,
1996b, 1997 for more detail and full references). The structure of the paper
is as follows.  In Section 2 we derive the upper limit on the binary
stochastic background in our Galaxy at frequencies 1 mHz - 0.1 Hz
(LISA diapason). At these frequencies, the background is mostly due to
binary WD mergers. We use the SN Ia statistics to constrain
galactic binary WD merging rate $\R <1/300$ yr$^{-1}$.  The
upper limit on the background at these frequencies is
$h_{lim}(f)\simeq 4 \times 10^{-20} (f/10^{-3} \hbox{Hz})^{-2/3}$ for
the inverse-average distance to typical binary WD 10 kpc. In Section
3, we calculate the binary galactic background using Monte-Carlo
modeling of binary star evolution in the Galaxy. The comparison of the
calculated backgrounds with the proposed LISA sensitivity is given in
Section 4. In Appendix A we give an alternative derivation of equation 
(\ref{h_c})
for the part of the background formed by coalescing binary stars only.
In Appendix B we briefly discuss cosmological effects. 

\section{Upper bounds on the binary stochastic
GW background}

It is widely recognized that merging compact binary stars
(WD+WD, NS+NS ...) determine the high-frequency part of
binary stochastic background (e.g., Lipunov, Postnov \& 
Prokhorov 1987; Hils, Bender \& Webbink 1990). In this
Section we wish to show that this portion is mostly
significant in the LISA frequency band and provides an upper
limit  on this background in general. 

\subsection{GW backgrounds formed by 
merging binary stars}

Consider a model galaxy consisting entirely of binary
systems.  Let us assume a stationary star formation rate (which
describes well the situation in our Galaxy).  We
will be interested in only LISA frequency range, $10^{-4}-10^{-1}$ Hz,
inside which only coalescing binary white dwarfs and binary neutron
stars contribute. Even if binary neutron stars coalesce at a rate of
1/10000 yr in the Galaxy (Lipunov et al. 1987; 1995; Tutukov and
Yungelson 1993), their number still should be much smaller than the white
dwarf binaries, and in this section we restrict ourselves to
considering only binary WD.

As mentioned in the Introduction, the stochastic GWB may be fully
characterized by the energy emitted per logarithmic frequency
interval. In the case of a binary system this energy is exactly equal
to the change in the orbital energy of two stars during the time
period needed for the GW frequency (which is twice the orbital
frequency for circular orbits) to pass the frequency interval $\Delta
f\approx f$. Therefore, in the stationary situation the net energy
emitted in GW will be determined by the rate $\R $ at which the systems
enter the specific frequency interval and the rate of their orbital
frequency change $\dot f$. Clearly, if  the binary orbit evolves
only due to GW emission, the result will depend on the rate $\R $
only.

Indeed, the stationarity implies that the number
of binary WD per unit logarithmic frequency interval
may be determined from the continuity equation
\begin{equation}
dN/d\ln f \equiv N(f) = \R \times  (f/\dot f)\,.
\label{contin}
\end{equation}
The total energy emitted per second per unit logarithmic frequency
interval at $f$ by all such binaries in the galaxy is
\begin{equation}
dE/(dt\,d\ln f) \equiv
L(f)=\sum_i L_i(f) = \widetilde L(f) N(f) = \widetilde L(f) \R \times (f/\dot f)\,,
\end{equation}
where $\widetilde L(f)\propto (\widetilde \M f)^{10/3}$ 
is the characterictic GW luminosity at
frequency $f$ which is dependent on the so-called ``chirp mass''
${\M}$ of the binary system:
\begin{equation}
{\cal M }= M(\mu/M)^{3/5}
\end{equation}
($M$ is the total and $\mu$ reduced masses).
We use $\widetilde {\M}$ for some average mass
of the typical binary (see discussion followed eq. [\ref{omega}] below). 
For the orbital frequency change due to GW we find
\begin{equation}
(f/\dot f)_{GW}=(2/3)(E_{orb}/\dot E_{orb})_{GW}
\end{equation} 
(we have used the fact that $E_{orb}\propto M_1M_2/a$ and
the third Kepler's law for binary semimajor axis $a$),
so if $\dot E_{orb}=(dE/dt)_{GW}$ we arrive at 
\begin{equation}
L(f)=(2/3)\R  E_{orb}(\widetilde {\M},f) \,.
\end{equation}

For an isotropic background we have 
\begin{equation}
\Omega_{GW}(f)\rho_{cr} c^2 = L(f)/(4\pi c \langle r \rangle^2)
= \R  E_{orb}(\widetilde {\M},f)/(6\pi c \langle r
\rangle^2)\,,
\label{omega}
\end{equation}
where $\langle r \rangle$ is
the inverse-square average distance to the typical source.
Strictly speaking, this distance (as well as 
the chirp mass) may be a function of frequency
since the binaries characterized by different 
chirp masses $\cal M$ may be differently distributed
in the galaxy. We are highly ignorant about the real distribution
of binaries in the galaxy, but taking the mean photometric 
distance for a spheroidal distribution in
the form 
$$dN\propto \exp[-r/r_0]\,\exp[-(z/z_0)^2]$$ 
($r$ is the radial
distance to the galactic center and $z$ the hight above the galactic
plane) with $r_0=5$ kpc and $z_0=4.2$ kpc with  
$\langle r \rangle \approx 7.89$ kpc is sufficient for our purposes.

Substituting $E_{orb}\sim {\M}c^2({\M}f)^{2/3}$ into equation
(\ref{omega}) we obtain
\begin{equation}
\Omega_{GW}(f)\approx 2\times 10^{-8} \R_{100} 
(f/10^{-3} \hbox{Hz})^{2/3}(\widetilde {\M}/M_\odot)^{5/3}( \langle r \rangle/10\,
\hbox{kpc})^{-2}h_{100}^{-2}\,,
\end{equation}
where $\R_{100}=\R /(0.01$ ~yr$^{-1})$ is the galactic
rate of binary WD mergers.

In terms of the characteristic dimensionless amplitude
of the noise background that determines the signal-to-noise
ratio when cross-correlating outputs of two
independent interferometers (cf. Thorne 1987, eq. [65])
we have
\begin{equation}
\eqalignleft{
h_c(f)&=(1/2\pi)(H_0/f)\Omega_{GW}^{1/2}   \cr
      &\approx 7.5 \times 10^{-20} \R_{100}^{1/2} 
(f/10^{-3} \hbox{Hz})^{-2/3}(\widetilde {\M}/M_\odot)^{5/6}
( \langle r \rangle/10\,\hbox{kpc})^{-1}
\cr}
\label{h_c}
\end{equation}
irrespective of $H_0$ (naturally). Explicit derivation of
equation  (\ref{h_c}) is given in the Appendix A.
Equation  (\ref{h_c}) shows that at high frequencies of interest here
the GW background is fully determined by the galactic rate of binary WD
mergers and is independent of (complicated) details of binary
evolution at lower frequencies (the examples of calculated spectra at
lower frequencies see in Lipunov \& Postnov 1987; Lipunov, Postnov \&
Prokhorov 1987; Hils et al. 1990, and below).

All above considerations are valid for frequencies at which more than 
one binary system fall within the logarithmic
frequency interval $\Delta f\approx f$. Formally this limiting frequency 
is specified by the requirement $N(f)\ge 1$, i.e.
\begin{equation}
f<f_{cr}=0.1\hbox{Hz}\,(\widetilde {\M}/M_\odot)^{-5/8}\R_{100}^{3/8}\,,
\label{f_cr}
\end{equation}
which is, however, already close to the limiting orbital frequency for
a double WD binary.  As the number of systems per logarithmic
frequency interval rapidly decreases with frequency, $N(f)\propto
f^{-8/3}$, the stochastic background starts forming shortly below this
limiting frequency. Note that combining equations (\ref{h_c}) and
(\ref{f_cr}) we can find the relationship $h(f_{cr}) \ge 8\times
10^{-22} (f_{cr}/0.01 \hbox{Hz})^{2/3} (\M/M_\odot)^{5/3}(r/10
\hbox{kpc})^{-1}$ above which binary-merger-formed GW background
appears; this limit depends on the mean galactic distance $r$, which
is known much more reliably than binary WD merger rate $\R$. For $\M<1
M_\odot$, however, this boundary lies below the rms LISA sensitivity
at all frequencies.

\subsection{Notes on the extragalactic binary GW background}

The derivation of the binary background given above (see also
Appendix A) becomes more accurate for extragalactic binaries. The
contribution from extragalactic binaries may be shown to be smaller 
than from galactic ones at
all frequencies (see, e.g. Hils et al. 1990, Lipunov et al. 1995), but
for completeness we give here the final result in terms of $\Omega_{GW}$
which bears a clear physical meaning.

%) and refer to Appendix B for
%more detail.

At the detector's frequency $f=f'/(1+z)$, galaxies
lying inside the proper volume $dV(z)$ at redshift $z$ contribute
\begin{equation}
\rho_{cr}c^2 d\Omega_{GW}(f)=L(f')/(4\pi c D(z)^2)n(z) dV(z) 
\label{dO}
\end{equation}
where $L(f)$ is the GW luminosity at the source frequency $f$
within each galaxy, $D(z)$ is the luminosity distance, $n_G$ is the 
present density of galaxies and $n(z)=n_G(1+z)^3$
is the density of galaxies at the redshift $z$ without
evolutionary effects.
In the Universe with zero curvature $dV(z)=4\pi d^2(z) d(d(z))$, 
where $d(z)=D(z)/(1+z)$ is the proper motion distance. 
Taking into account that the present density of 
galaxies can be rewritten through the fraction of
luminous matter baryons,

\begin{equation}
n_G \approx 0.013( \hbox{Mpc}^{-3})(\Omega_b/0.005)h_{100}^2\,. 
\end{equation}
equation (\ref{dO}) can be recasted in the form 
\begin{equation}
d\Omega_{GW}(f) = (2/3)\Omega_b{\cal R}\,(E_{orb}(\M,f(1+z))/M_G c^2)\,
(1+z)d(d(z))/c
\label{dO(z)}
\end{equation}
where $M_G$ is the baryon mass of a typical galaxy per which the
merging rate $\R$ is calculated.

For flat ($\Omega=1$) standard cosmological model without
$\Lambda$-term $d(z)=(2c/H_0)(1-1/\sqrt{1+z})$. Noticing
that $E_{orb}(f(1+z))$ scales as $(1+z)^{2/3}$ 
(eq. [\ref{eorb_a}] from Appendix A), we
may integrate equation (\ref{dO(z)}) up to $z_*$ -- the redshift
at which  first WD mergers had started -- to obtain
\begin{equation}
\eqalignleft{ 
\Omega_{GW}(<z_*)&=(6/7)\Omega_b t_H\R (E_{orb}/M_Gc^2)
((1+z_*)^{7/6}-1)\cr 
&= (3/7)\Omega_b t_H\R (\M/M_G)
(x/2)^{2/3}((1+z_*)^{7/6}-1)\cr 
&\approx 10^{-9}
(\Omega_b/0.005)\R_{100}\cr &\qquad(10^{11}\M/M_G)(\M/M_\odot)^{2/3}
(f/1\hbox{Hz})^{2/3}h_{100}^{-1}((1+z_*)^{7/6}-1)\,.  \cr}
\label{O(<z)}
\end{equation}
This expression bears very clear physical meaning simply as the fraction
of energy emitted in GW over the Hubble time ($t_H=2/3H_0$) by
WD mergers with respect to the rest-mass energy of baryons in stars
in the Universe.  The ratio between the cosmological and galactic backgrounds
is
\begin{equation}
\eqalignleft{
\Omega_{GW}(z_*)/\Omega_{GW}& = 
\langle r \rangle^2\int^{z_*}(1+z)^{2/3}n(z)dV(z)/D(z)^2\cr 
&\approx 0.05 (\Omega_b/0.005)(\langle r\rangle/10\,\hbox{kpc})^2
h_{100}^{-1}((1+z_*)^{7/6}-1)
\cr}
\end{equation}
regardless of the poorly known merger rates, which shows that
cosmological binary background is generally a few times smaller than the
galactic one.  Strong source evolution with redshift would increase
this ratio, but at present we cannot calculate reliably these effects.

\subsection{SN Ia rate as the upper limit of binary WD merger rate}

The galactic merger rate of close binary WD is unknown.
One possible way to recover it is  searching for close white
dwarf binaries. A recent study (Marsh et al. 1995),
revealed a larger fraction of such systems than had previously been
thought. Still, the statistics of such binaries in the Galaxy remains
very poor.

If coalescing binary WD are associated with SN Ia explosions, as
proposed by Iben \& Tutukov (1984) and further investigated by many
authors (for a recent review of SN Ia progenitors see Branch et
al. 1995), their coalescence rate can be constrained using much more
representative SN Ia statistics. Branch et al. (1995) concluded that
coalescing CO-CO binary WD remain the most plausible candidates mostly
contributing to the SN Ia explosions.  The galactic rate of SN Ia is
estimated $4\times 10^{-3}$ per year (Tamman et al.  1994; van den
Bergh and McClure 1994), which is close to the calculated rate of
CO-CO coalescences ($\sim (1-3)\times 10^{-3}$). The coalescence rate
for He-CO WD and He-He WD (other possible progenitors of SN Ia) falls
ten times short of that for CO-CO WD (Branch et al. 1995).  

We note further that binary WD mergers as main progenitors of the SN
Ia explosions start falling out of favour in the last years (see
especially critical studies by Nomoto et al. 1996). Evolutionary,
double WD are formed through the common envelope stage (Iben and
Tutukov 1984, Webbink 1984).  In 1990s, with the calculation of new
opacities, the possibility appeared for a WD, accreting even at a very
high rate from the secondary companion in a binary system, 
to avoid the common envelope
formation because of efficient stellar wind (Hachisu et al. 1996).
This mechanism decreases the double WD formation rate and explains why the
observed space density of binary degenerate dwarfs is smaller than
derived from Iben and Tutukov's scenario. Therefore, we may conclude
that the observed SN Ia rate provides a secure {\it upper limit} to
the double WD merger rate regardless of the evolutionary
considerations.

\section{A model of the galactic binary  GW background}

In this section we discuss the galactic GW background calculated with
the Scenario Machine code for binary population synthesis in our
Galaxy (see Lipunov, Postnov \& Prokhorov 1995, 1996b for full
description of the code). Briefly, we use a model galaxy with the
total stellar mass of $10^{11}$ M$_\odot$ and half mass in
binaries. We assume a constant star formation rate of 1 M$_\odot$ per
year which is a reasonable approximation for our Galaxy. Assuming the
Salpeter mass distribution function $dN/dM \propto
(M/M_\odot)^{-2.35}$ , this star formation rate requires the minimal
mass of the star to be $M_{min}=0.1$ M$_\odot$ in order to produce the
total stellar mass of the Galaxy during the Hubble time of 15 billion
years. 
%The maximum mass of the star to form is assumed to be 120
%M$_\odot$, which in fact is not important. 
The initial mass ratios of
the modelled binaries $q=M_2/M_1$ are assumed to be uniformly
distributed from 0 to 1 (see, however, discussion in Lipunov, Postnov
\& Prokhorov 1996a). The important evolutionary parameter -- the
efficiency of the common envelope stage $\alpha_{CE}$ -- was fixed at
1.  Another crucial evolutionary parameter, the distribution of the
kick velocity imparted to a neutron star at birth, is not very
important here as we are interested mostly in low-mass stars (with
initial masses $M_1\le 10$ M$_\odot$) which evolve ultimately into
WDs.

The position of the Sun in the outskirts of the Galaxy makes the
binary stochastic background anisotropic (see Lipunov et
al. 1995). The exact calculations of this background thus would require
the knowledege of the spatial distribution of binary stars in the Galaxy,
which is uncertain, so we restrict ourselves to consider the average
amplitude of the stochastic background using the mean photometric
distance in the Galaxy 7.9 kpc as explained in the previous section.

The calculated background is shown in Fig. 1 by the thick solid line.
For comparison, we plotted Bender et al's data (filled quadrangles).

\placefigure{fig1}
%[FIG. 1 GOES HERE]

\section{Discussion and conclusions} 

The upper frequency  at which the stochastic GW background formed by galactic
merging binaries becomes smaller than the detector's noise limit  $h_{rms}(f)$ 
may be derived from equation  (\ref{h_c}):
$$
f_{lim}\approx (0.02 \hbox{ Hz})\times (h_{rms}/10^{-20})^{-3/2}\R_{100}^3
(\M/M_\odot)^{5/4}(r/10 \hbox{ kpc})^{-3/2} 
$$

The upper limit (\ref{h_c}) is plotted in Fig. 1 for different rates
of binary WD mergers $\R_{100}=1, 1/3, 1/10, 1/30$ assuming the chirp
mass ${\M}\approx 0.52 M_\odot$ (as for two CO white dwarfs with equal
masses $M_1=M_2=0.6 M_\odot$).  These lines intersect the proposed
LISA rms sensitivity at 

\begin{equation}
f>f_{lim}\approx 0.03-0.07 \hbox{Hz}\,.
\label{lim}
\end{equation}

This means that at frequencies higher than $0.07$ Hz no continuous
GW backgrounds of galactic origin are presently known to contribute 
above the rms-level
of LISA space laser interferometer.  The contribution from
extragalactic binaries is still lower regardless of
the poorly known binary WD merging rate (at least 
in the limit of no strong source evolution with $z$). Other
possible sources could be extragalactic massive BH binary systems
(e.g. Hils and Bender 1995). Their number in the Universe can be fairly
high (e.g. Rees 1997), but no reliable estimates of their contribution
are available at present. The lower limit (\ref{lim}) is already close
to the LISA sensitivity limit at 0.1 Hz, but we stress that the
assumptions used in its derivation are upper limits, so the actual
frequency beyond which no binary stochastic backgrounds contribute may
be three times lower. This precise limit depends on the details of binary
WD formation and evolution which are still poorly  known.

Fig. 1 demonstrates that the calculated GW background intersects LISA
sensitivity curve at frequencies $\sim 0.05$ Hz, and Bender et al's
curve at even lower frequencies $\sim 0.01$ Hz. The latter is
probably due to Bender et al's curve being derived from observational
estimate of double WD galactic density in the solar neighborhoods; we
stress once more that once formed, the binary WD will evolve until the
less massive companion fills its Roche lobe; unless the mass ratio is
sufficiently far from one (cf. Webbink 1984), the merger should
occur. Therefore Bender et al's curve provides a secure {\it lower
limit} to the galactic binary stochastic GW background.  We also note
that if the coalescence of two WD with inequal masses is prevented by mass
transfer process (then the orbital period of the system begins
increasing), some features on the shape of the background may emerge
(cf. discussion of the minimum orbital period for cataclysmic
variables in Lipunov, Postnov \& Prokhorov 1987).

Presently, we cannot rule out the high galactic double WD merger rate
(1/300-1/1000 yr$^{-1}$), and therefore can consider $f_{lim}$ to lie
within the frequency range $0.01-0.07$ Hz.  We conclude that no GW
background of galactic origin above this frequencies should contribute
at the rms-noise level of LISA interferometer, and hence the detection
of an isotropic stochastic signal at frequencies $0.03-0.1$ Hz with an
appreciable signal-to-noise level (which possibly may be done using
one interferometer) would strongly indicate its cosmological origin.
To be detectable by LISA, the power of relic GW background should be
$\Omega_{GW}h_{100}^2>10^{-8}$ in this frequency range.

\acknowledgments

The work was partially supported by Russian Fund for Basic Research
through Grant No 95-02-06053-a and by INTAS Grant No 93-3364. 
We thank Prof. Leonid Grishchuk for stimulating
discussions. KAP acknowledge the Department of Physics and Astronomy
of University of Wales (Cardiff) for hospitality and financial support
of his visit to Cardiff. Important part of this wark was done during
visit of KAP to the Cosmic Radiation Laboratory of RIKEN (Japan),
which was made possible through STA Fellowship Grant No 496057.

\appendix
\section{Explicit derivation of equation (\protect\ref{h_c})} 

Equation (\ref{h_c}) can be derived explicitly, 
without using the notion of $\Omega_{GW}$ (instead, 
one can  use $h_c$ to determine $\Omega_{GW}$;
both means are, of course, fully equivalent).
We reproduce  here this equation keeping
proportionalities
$$
h_c(f)\propto (1/r)\R^{1/2}\M^{5/6}f^{-2/3}\,.
$$
We find it convenient to introduce new variables:

$$
\eqalignleft{
&R_g=2G\M/c^2, \qquad\hbox{gravitational radius of the chirp mass}\,,\cr
&R_l=c/(\pi f), \qquad\hbox{light cylinder radius for orbital
frequency}\,,\pi f/2
\cr}
$$
and their dimensionless ratio 

$$ 
x(\M, f)\equiv x = (R_g/R_l)\propto \M f\,.
$$
This units allows us to write all relevant quantities in a compact form
with clear physical meaning keeping $G$ and $c$ and facilitating
numerical estimates.
In these units, the orbital energy of a binary system with chirp
mass $\M$ is

\begin{equation}
E_{orb}=-(1/2)GM_1M_2/a=-(1/2)\M c^2 (x/2)^{2/3}\,.
\label{eorb_a}
\end{equation}

The energy flux per unit time per unit area
carried by a GW is
\begin{equation}
dE/(dAdt)=c^3 /(16\pi G)\overline{ (\dot h_+(t)^2+\dot h_\times(t)^2)}
\label{flux}
\end{equation}
where $h_+$ and $h_\times$ are wave polarizations, the overbar
means averaging over several cycles of the wave. 
Switching to the frequency domain in usual way (see, e.g., Thorne
1987), we obtain

\begin{equation}
dE/(r^2 d\Omega df) =(c^3/G) (\pi f^2/2) (|\tilde h_+(f)|^2+|\tilde
h_\times (f)|^2)\,,
\label{spektr}
\end{equation}
where $r$ is the distance to the source, $d\Omega=dA/r^2$ 
is the elementary solid angle.

Let us describe the stochastic backgrounds formed by 
many independent sources within unit logarithmic
frequency interval at $f$
by the characteristic strain amplitude defined as
(cf. eq. [\ref{flux}])
\begin{equation}
h_c^2(f)=4G/(\pi c^3) (1/f^2) \sum_i^{N(f)}dE_i/(r_i^2d\Omega dt)
\label{h_c'}
\end{equation}
where $N(f)=\R f/\dot f$ is the number of sources assuming the
stationary source birth rate $\R$ (cf. eq. [\ref{contin}]). We wish to
show that this equation gives exactly the same result as equation
(\ref{h_c}).

Substituting equations (\ref{flux}) and (\ref{spektr}) into 
equation (\ref{h_c'}),
using equation (\ref{contin}), and introducing, as before, 
some effective distance 
and the chirp mass of a typical source (now we will not
mark them by special symbols), we arrive at 
\begin{equation}
h_c^2(f)=\R f\langle |\tilde h(f)|^2\rangle \,,
\label{h_c''}
\end{equation}
where $\langle |\tilde h(f)|^2\rangle =
 \langle\sum_{i=1}^2 |\tilde h_i(f)|^2\rangle $ stands for 
angle-averaged Fourier-components (with dimension [Hz$^{-2}$]) of $h(t)$:
\begin{equation}
\eqalignleft{
\langle |\tilde h(f)|^2\rangle &=
(1/4\pi)(1/f^2)\int 2G/(\pi c^3)(dE/r^2dfd\Omega) d\Omega\cr
&=(1/4\pi r^2) 2G/(\pi c^3)(dE/df)/f^2 \,.
\cr}
\label{fourier}
\end{equation}
(Note that eq. [\ref{h_c''}] can also be derived directly 
from Parseval's theorem and the expression for the 
variance per logarithmic frequency
interval of a stochastic noise associated with 
unresolved  independent sources).

Now notice that for circular orbits (our initial assumption)
$dE/df=(dE_{orb}/df)$ if far from coalescence (which is the case
under consideration), so 
\begin{equation}
dE/df= - (1/3) (\M c^2/f) (x/2)^{2/3} \propto \M^{5/3}f^{-1/3}\,,
\end{equation}
and from equation (\ref{fourier})
\begin{equation}
\langle |\tilde h(f)|^2\rangle=(1/12\pi r^2)R_l^2/f^2 (x/2)^{5/3}
\propto \M^{5/3}f^{-7/3}
\end{equation}
(cf. eq. [44] in Thorne 1987). Finally, from equation (\ref{h_c''}) we obtain
\begin{equation}
h_c^2(f)=(\R/6\pi)(R_g/r)^2 (x/2)^{-1/3}/f \propto \M^{5/3}f^{-4/3}\,,
\end{equation}
which yields exactly equation  (\ref{h_c}).

\clearpage

\figcaption[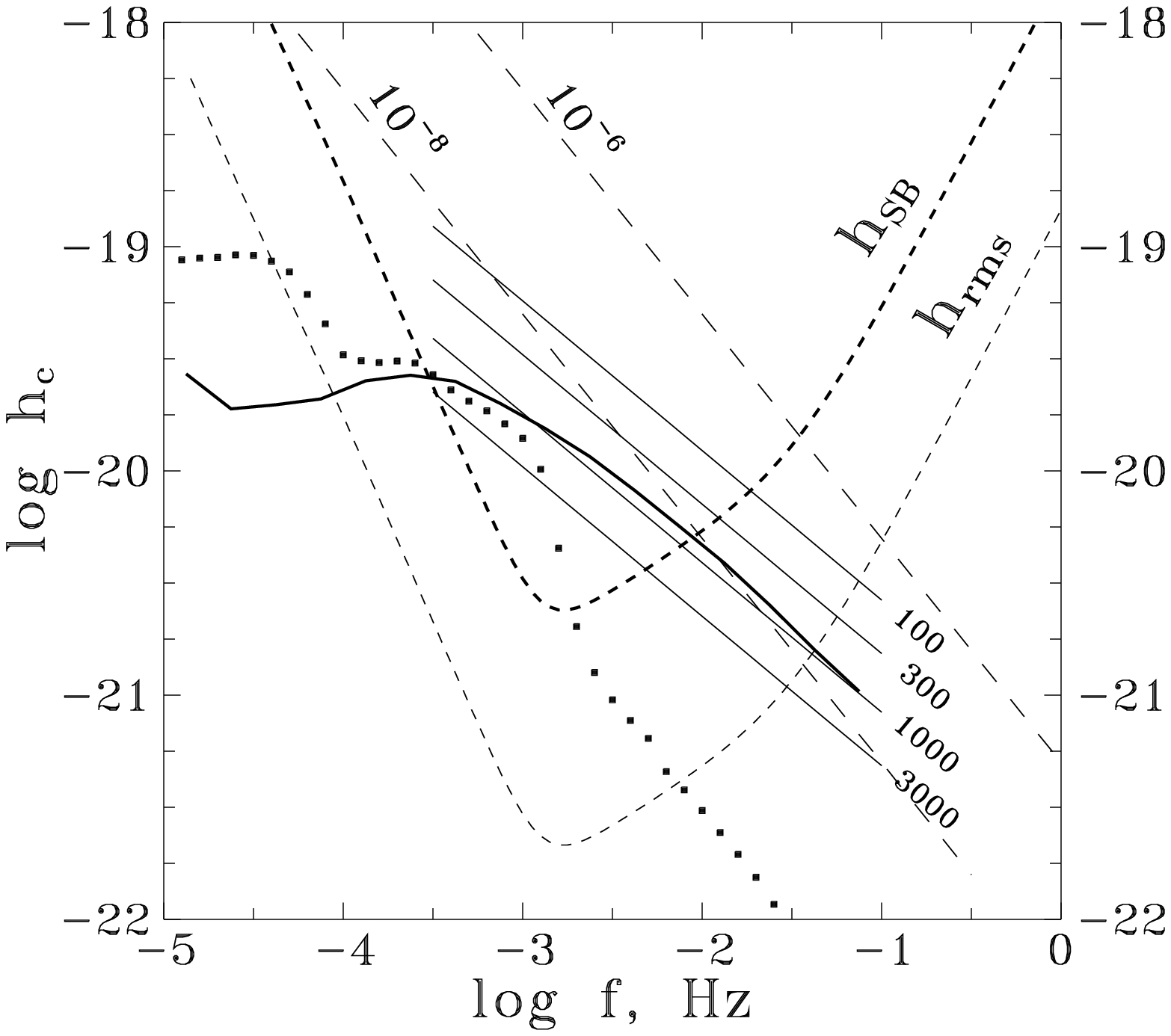]{
Galactic binary GW background $h_c$ as given by Bender (1996)
(filled quadrangles) and calculated for a model spiral galaxy
with the total stellar mass $10^{11}$ M$_\odot$ (the solid curve).
Average photometric distance 7.9 kpc
is assumed. Thin straight lines marked with 100, 300, 1000, 3000 are the
analytical upper limit (eq. [\protect\ref{h_c}]) for binary WD merger rates
1/100, 1/300, 1/1000, and 1/3000 yr$^{-1}$ in a model spiral galaxy,
respectively, assuming \protect{$\M=0.52$} M$_\odot$.  
Straight dashed lines labeled by $10^{-8}$, $10^{-6}$ show GW
backgrounds corresponding to constant $\Omega_{GW}$. The proposed LISA rms
noise level ($h_{rms}$) and sensitivity to bursts
$h_{SB}=5\protect\sqrt{5} h_{rms}$ are also reproduced (cf. Thorne 1995;
Fig. 14).\label{fig1}}

\end{document}